\documentclass[twocolumn,description,aps]{revtex4-1}
\usepackage{graphicx,epstopdf,amsmath,amssymb,multirow,CJK,color,tabularx}
\usepackage[german,english]{babel}
\usepackage[colorlinks, linkcolor=blue,anchorcolor=blue,citecolor=blue,urlcolor=blue]{hyperref}

\begin{document}
\title{Ideal Nodal-Sphere Semimetal in the Three-Dimensional Boron Allotrope CT-B$_{24}$}

\author{Xiao-jing Gao$^{1}$}
\author{Yanfeng Ge$^{1}$}
\author{Yan Gao$^{1}$}\email{yangao9419@ysu.edu.cn}

\affiliation{$^{1}$State Key Laboratory of Metastable Materials Science and Technology $\&$ Hebei Key Laboratory of Microstructural Material Physics, School of Science, Yanshan University, Qinhuangdao 066004, China}

\date{\today}

\begin{abstract}
Nodal-sphere semimetals (NSSMs), featuring spherical band degeneracies in momentum space, constitute a fascinating class of topological materials. However, their realization in real materials is severely hampered by discrete crystallographic symmetry constraints, often resulting in gapped ``pseudo'' nodal spheres. Here, combining first-principles calculations and symmetry analysis, we predict a new three-dimensional boron allotrope, CT-B$_{24}$, as a nearly ideal NSSM. Its structural stability is systematically confirmed by phonon calculations, \textit{ab initio} molecular dynamics simulations at 600~K, and elastic constant analysis. Notably, the electronic structure of CT-B$_{24}$ exhibits two bands crossing linearly near the Fermi level, forming a quasi-nodal sphere around the $\Gamma$ point. The maximum energy gap is merely 0.008~meV, which is two orders of magnitude smaller than the gaps reported in previous pseudo-NSSMs. Furthermore, the (001) surface hosts pronounced drumhead-like surface states located outside the projected nodal sphere, providing distinct signatures detectable by angle-resolved photoemission spectroscopy (ARPES). The nodal sphere also demonstrates remarkable robustness and tunability under external strain, driving a topological phase transition from an NSSM to a Dirac semimetal and finally to a trivial insulator. Our work not only presents a superior material platform for exploring nodal-sphere physics but also suggests potential for strain-tunable topological devices.
\end{abstract}

\date{\today} \maketitle

\section{INTRODUCTION}\label{sec_introduction}

Topological semimetals (TSMs) have attracted considerable attention due to the unique properties and potential applications of the emergent quasiparticles that arise from their band crossings near the Fermi level in momentum space~\cite{WDSMs,TSMs,TSMsFab,Yu-SB}. The dimensionality of these band degeneracies provides a natural classification scheme for TSMs: zero-dimensional (0D) Weyl~\cite{WanWSM,WengTaAs}, Dirac~\cite{YoungDSM,WangNa3Bi}, and three-, six-, and eightfold degenerate points~\cite{UnMPBradlyn,TPs}, one-dimensional (1D) nodal lines~\cite{FangNLSM,GaoNLSM,MTCsNLSM}, and the recently proposed two-dimensional (2D) nodal-surface semimetals~\cite{YangNS,ZhongNSs,ChenNS} and nodal-sphere semimetals (NSSMs)~\cite{MorozWNS,PDNS}. Among the various TSM phases, NSSMs occupy a unique position in the hierarchy of band crossings. In a NSSM, the conduction and valence bands would touch along an entire spherical surface near the Fermi level in momentum space, creating a closed 2D degeneracy manifold. Such systems are predicted to exhibit extraordinary physical properties~\cite{PDNS,UribeSHE}, including direction-independent drumhead surface states, constant density of states near the Fermi level, and localized spin Hall conductivity arising from the spherical Fermi surface topology. These characteristics distinguish NSSMs from their lower-dimensional counterparts and suggest potential applications in quantum devices and low-energy electronics.

However, the realization of ideal nodal-sphere states in crystalline materials is fundamentally hindered by the discrete nature of crystallographic symmetries~\cite{PDNS}. Band degeneracies in crystals are typically confined to high-symmetry lines or planes, where symmetry eigenvalues enforce protected crossings. In contrast, a spherical degeneracy would require symmetry-mandated crossings at arbitrary momentum points, a condition generally incompatible with the symmetry constraints of real crystal lattices. To circumvent this limitation, the concept of a pseudo-Dirac nodal sphere (PDNS) semimetal has been proposed. In such systems, symmetry-protected nodal rings on high-symmetry planes form a 3D skeleton, while the regions between them remain quasi-degenerate, often protected by higher-order tensor terms (HOTTs)~\cite{PDNS}, resulting in a small but finite gap. When HOTTs are sufficiently weak, the system closely approximates an ideal nodal sphere. To date, several PDNS candidates have been predicted with varying gap sizes~\cite{PDNS,UribeSHE,SiNS}, including LaN and CaTe ($\Delta E \sim 2$ meV), Tl$_5$Se$_2$Br and its analogs ($\Delta E \sim 0.5$ meV), I4$_1$/acd-Si ($\Delta E \sim 1$ meV), as well as MH$_3$ (M = Y, Ho, Tb, Nd) and Si$_3$N$_2$, both with gaps on the order of meV. Notably, most known PDNS materials contain heavy elements, where spin-orbit coupling (SOC) can substantially increase the gap and destabilize the topological phase. In contrast, light-element systems such as boron and carbon, with negligible SOC, offer a promising platform for realizing stable, near-gapless nodal-sphere states.

\begin{figure*}[!th]
	\centering
	\includegraphics[width=0.72\textwidth]{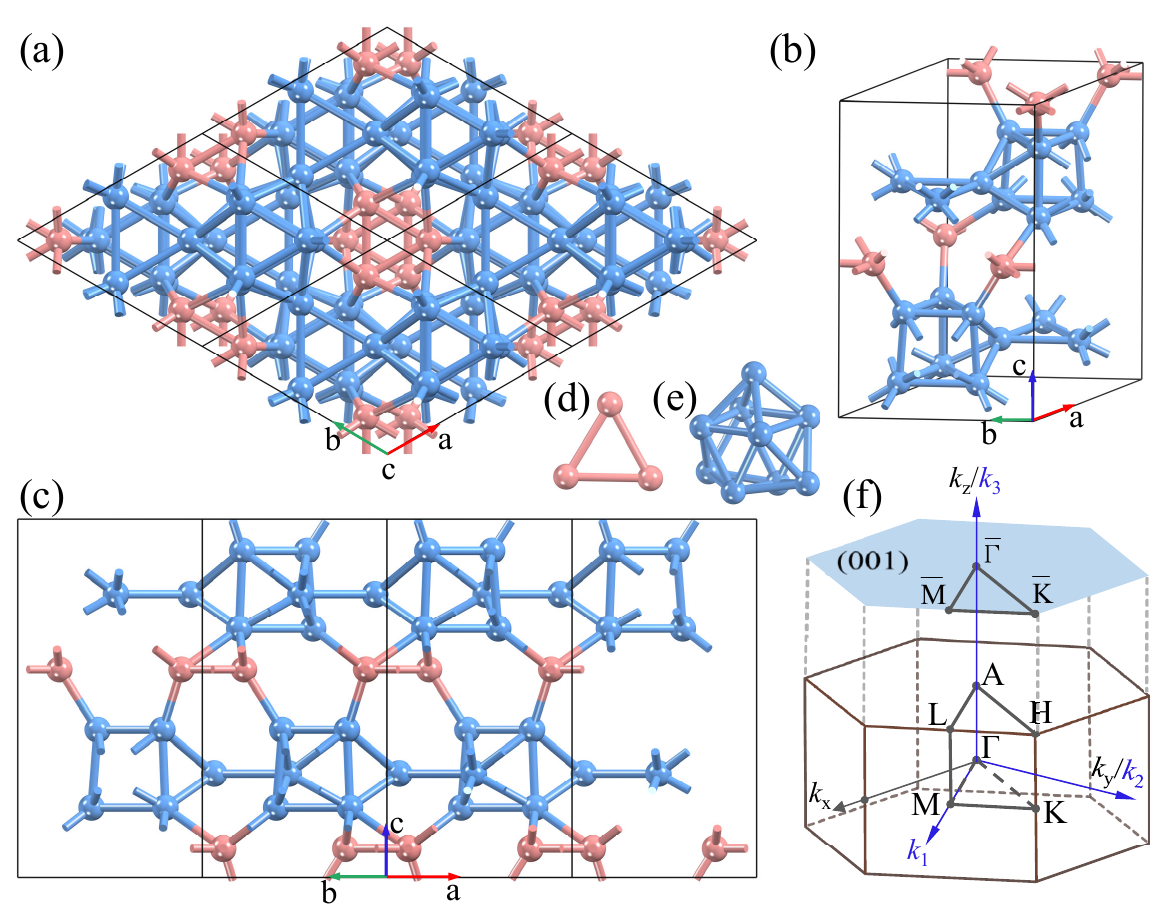}
	\caption{(Color online) Crystal structure and Brillouin zone of CT-B$_{24}$. (a) Top view and (c) side view of a $2\times2\times1$ supercell. (b) Perspective view of the primitive cell. (d) B$_{3}$ triangle and (e) B$_{9}$ cage building blocks. (f) The bulk Brillouin zone (BZ) and the projected (001) surface BZ.}
	\label{fig_structure}
\end{figure*}

Boron, the only nonmetal in group IIIA, exhibits unique electron-deficient bonding that supports multicenter interactions and a rich structural polymorphism~\cite{WangReBoron,Tai2DB,YakobsonBoron,Goddard3Dboron}, encompassing 0D clusters, 1D nanotubes, 2D borophenes, and three-dimensional (3D) polymorphs. In the context of 3D topological boron, several allotropes such as 3D borophene~\cite{3Dborophene}, 3D-$\alpha'$ boron~\cite{3DalphaB}, H-boron~\cite{H-boron}, Pnma-B$_{60}$~\cite{Pnma-B60}, AB-16-Pnnm~\cite{AB-16-Pnnm}, and Ort-B$_{32}$~\cite{Ort-B32} have been theoretically identified as nodal-line semimetals. Nevertheless, no boron allotrope hosting a nodal-sphere state has yet been reported. Given its negligible SOC and structural variety, boron is an ideal platform for realizing a nearly ideal PDNS semimetal, provided a suitable crystal structure can be identified.

\begin{table*}[!th]
	\renewcommand{\thetable}{I} 
	\caption{\label{tab:I} Structural parameters, mechanical properties, and electronic characteristics of CT-B$_{24}$ and selected boron allotropes, including space groups, lattice parameters (\AA), bond lengths (\AA), density (g/cm$^3$), total energy per atom (eV/B), bulk modulus (GPa), and electronic properties (NLSM: nodal-line semimetal; NSSM: nodal-sphere semimetal).}
	\centering
	\begin{tabular*}{2.0\columnwidth}{@{\extracolsep{\fill}}cccccccccc}
		\hline\hline
		\multirow{2}{*}{Structure} & \multirow{2}{*}{Space groups} & \multicolumn{3}{c}{Lattice parameters} & \multirow{2}{*}{Bond lengths} & \multirow{2}{*}{Density} & \multirow{2}{*}{$E_{\text{tot}}$} & \multirow{2}{*}{Bulk modulus} & \multirow{2}{*}{Properties} \\
		& & a & b & c & &  & & & \\ 
		\hline
		CT-B$_{24}$ & $P6_{3}mc$ & 4.99 & 4.99 & 8.39 & 1.66-2.02 & 2.37 & -6.47 & 172.63 & NSSM \\
		H-boron & $P6_{3}/mmc$ & 6.06 & 6.06 & 9.91 & 1.62-1.71 & 0.91 & -5.84 & 70.42 & NLSM \\
		Ort-B$_{32}$ & $Cmcm$ & 6.61 & 8.09 & 4.98 & 1.63-2.09 & 2.16 & -6.29 & 194.00 & NLSM \\
		3D borophene & $C2/m$ & 5.47 & 2.81 & 1.85 & 1.73-1.89 & 2.68 & -6.31 & 242.34 & NLSM \\
		3D $\alpha'$ boron & $Cmcm$ & 7.73 & 8.23 & 5.07 & 1.66-1.85 & 1.78 & -6.36 & 166.16 & NLSM \\
		AB-16-Pnnm & $Pnnm$ & 3.20 & 8.48 & 4.50 & 1.61-1.91 & 2.35 & -6.42 & 218.32 & NLSM \\
		Pnma-B$_{60}$ & $Pnma$ & 11.82 & 4.90 & 7.52 & 1.72-1.87 & 2.47 & -6.60 & 235.76 & NLSM \\
		$\alpha$-B$_{12}$ & $R\overline{3}m$ & 4.89 & 4.89 & 12.55 & 1.67-1.80 & 2.48 & -6.70 & 238.11 & Insulator \\
		\hline\hline
	\end{tabular*}
\end{table*}

In this work, we predict a new 3D boron allotrope, CT-B$_{24}$, which hosts an ideal PDNS state. Its structure comprises interconnected B$_9$ cage units and B$_3$ triangular motifs, exhibiting dynamic, thermal, and mechanical stability. Notably, our first-principles electronic structure calculations reveal that only two bands cross linearly near the Fermi level, forming a closed spherical degeneracy in momentum space with a minimal HOTTs-induced gap of $\approx$0.008~meV, which is the smallest value reported to date. Symmetry analysis shows that this PDNS is underpinned by a skeleton of six mirror-protected nodal rings arising from the C$_{6v}$ point group. Projection of the nodal sphere onto the (001) surface yields pronounced drumhead surface states.  Furthermore, the quasi-nodal sphere in CT-B$_{24}$ undergoes topological phase transitions under hydrostatic pressure or uniaxial strain along the $c$-axis, evolving from a nodal-sphere state to a Dirac semimetal and finally to a trivial insulator. Our results establish CT-B$_{24}$ as an exceptional platform for investigating nodal-sphere physics and developing topological quantum devices.

\begin{figure*}[!th]
	\centering
	\includegraphics[width=0.96\textwidth]{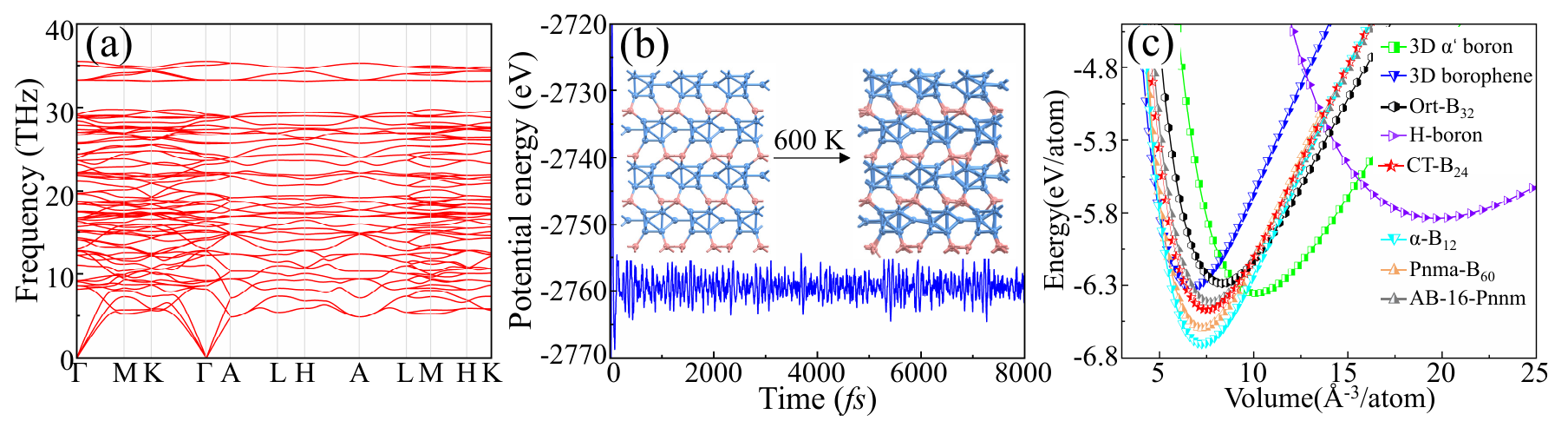}
	\caption{(Color online) Stability analysis of CT-B$_{24}$. (a) Phonon dispersion curves. (b) \textit{Ab initio} molecular dynamics (AIMD) simulation at 600~K: potential energy fluctuation over time (main panel) and structural snapshots before (left inset) and after (right inset) the 8000~fs simulation. (c) Energy-volume curves comparing CT-B$_{24}$ with other representative boron allotropes.}
	\label{fig_stability}
\end{figure*}

\section{METHOD}\label{sec_method}

We performed first-principles calculations within the density functional theory (DFT) framework using the VASP package~\cite{VASP} to investigate the structural, electronic, and topological properties of CT-B$_{24}$. The projector augmented-wave (PAW) method~\cite{PAW} was employed to describe the ion-electron interaction, and the exchange-correlation functional was treated with the Perdew-Burke-Ernzerhof (PBE) formalism~\cite{PBE} under the generalized gradient approximation (GGA). A plane-wave cutoff energy of 550~eV was used. Structural relaxation was carried out until the forces on all atoms were below 0.001~eV/\AA~and the total energy convergence reached $1\times10^{-6}$~eV. A Monkhorst-Pack $k$-point mesh~\cite{k-mesh} of $8\times8\times6$ was adopted for Brillouin zone (BZ) sampling during relaxation and electronic structure calculations. The phonon spectrum was computed with the finite displacement method using the Phonopy package~\cite{Phonopy} on a $2\times2\times2$ supercell. \textit{Ab initio} molecular dynamics (AIMD) simulations in the canonical ensemble with a Nos\'{e}-Hoover thermostat~\cite{AIMD} were conducted at 600~K for 8~ps to assess thermal stability. The elastic constants $C_{ij}$ were derived from energy-strain~\cite{elastic-constants} relations to examine mechanical stability. Topological properties and surface states were studied using maximally localized Wannier functions constructed with Wannier90~\cite{Wannier90} and analyzed with WannierTools~\cite{WannierTools}.

\section{RESULTS}\label{sec_results}

\subsection{Crystal Structure and Stability}
\label{subsec:crystal_stability}

The crystal structure of CT-B$_{24}$, a newly predicted 3D boron allotrope, adopts the hexagonal space group $P6_3mc$ (No.~186) with the corresponding $C_{6v}$ point group symmetry. Its primitive cell contains 24 boron atoms occupying four nonequivalent Wyckoff positions: 6c (0.468, 0.532, 0.210), 6c (0.225, 0.113, -0.078), 6c (0.567, 0.783, 0.589), and 6c (0.600, 0.800, 0.806), as illustrated in Figs.~\ref{fig_structure}(a) and~\ref{fig_structure}(b). The optimized lattice parameters are $a = b = 4.99$~\AA~and $c = 8.39$~\AA, yielding a mass density of 2.37~g/cm$^3$. The structure is composed of two distinct fundamental building blocks: triangular B$_3$ units and cage-like B$_9$ units [see Figs.~\ref{fig_structure}(d) and~\ref{fig_structure}(e)], which are interconnected through B-B covalent bonds. The B-B bond lengths range from 1.66 to 2.02~\AA, with bond angles varying between $43.29^\circ$ and $150.15^\circ$. The structural arrangement reveals an interesting stacking pattern along the $c$-axis, where identical units in adjacent layers are related by a $180^\circ$ rotation about the sixfold rotation axis [Fig.~\ref{fig_structure}(c)]. This stacking pattern, combined with the three vertical mirror planes ($\sigma_v$) and three diagonal mirror planes ($\sigma_d$) of the $C_{6v}$ point group, plays a critical role in stabilizing the topological PDNS state discussed below. The BZ and its (001) surface projection are shown in Fig.~\ref{fig_structure}(f).

\begin{figure*}[!th]
	\centering
	\includegraphics[width=0.96\textwidth]{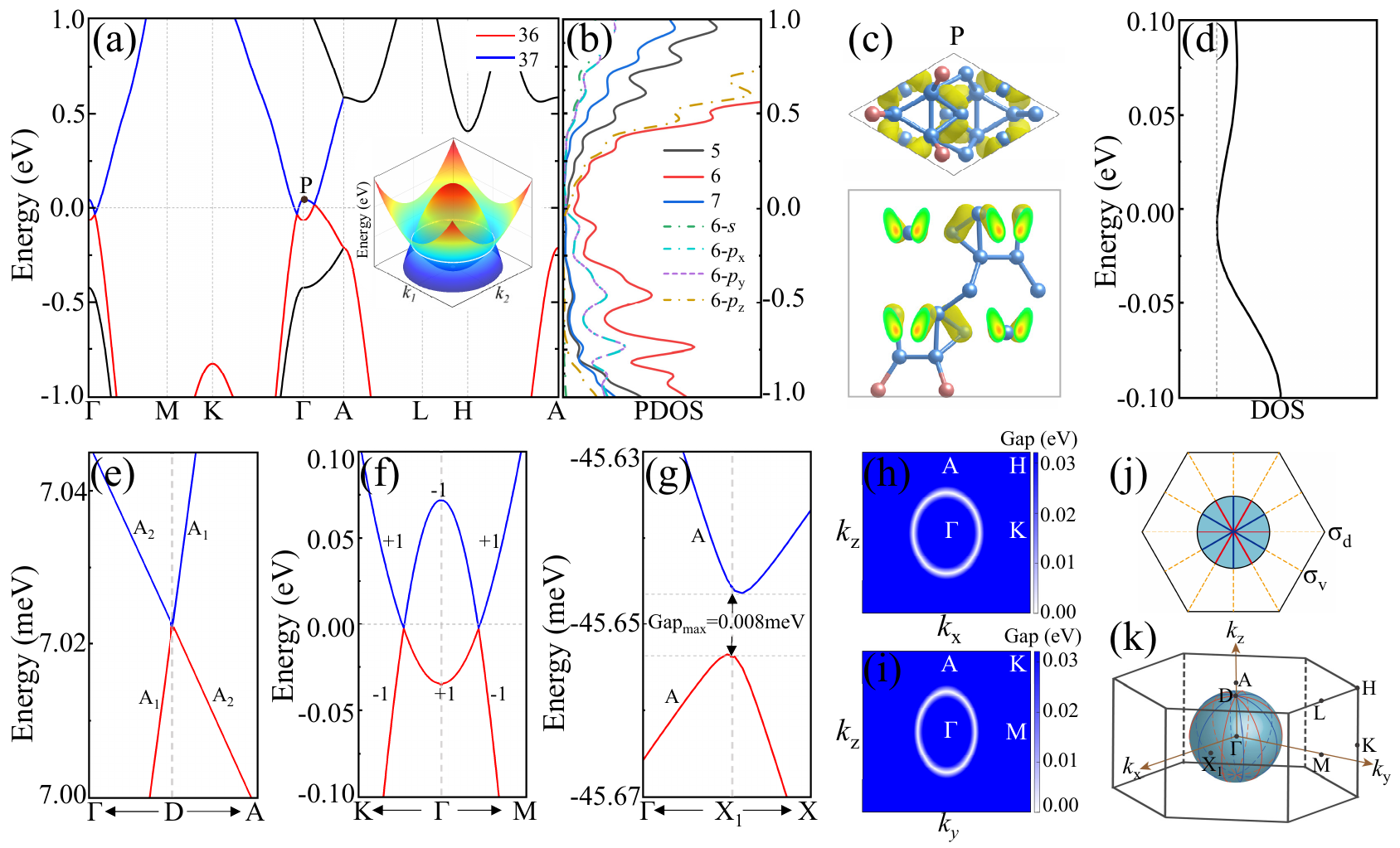}
	\caption{(Color online) Electronic and topological properties of CT-B$_{24}$. (a) Band structure along high-symmetry paths. Inset: 3D visualization of the band crossing in the $k_1$--$k_2$ plane. (b) Projected density of states (PDOS). (c) Top and side views of the charge density distribution at the P point labeled in Fig.~\ref{fig_structure}(a). (d) Density of states (DOS) near the Fermi level ($E_{\text{F}}$), with the dashed line indicating a constant value. (e) Band crossing at a gapless D point on the spherical surface. (f) Band structure around the $\Gamma$ point along the K-$\Gamma$-M path, showing the mirror eigenvalues of the crossing bands. (g) Band crossing at the $X_1$ point with a maximum gap of 0.008~meV, where $X_1$ is located on the spherical surface. Band gap maps on the (h) $k_x$--$k_z$ and (i) $k_y$--$k_z$ planes, with white curves highlighting the mirror-symmetry-protected nodal rings in the $C_{6v}$ point group. Top (j) and perspective (k) views of the quasi-nodal sphere in the BZ. Yellow dashed lines depict the three $\sigma_v$ and three $\sigma_d$ mirror planes of the $C_{6v}$ point group; blue and red solid curves represent the mirror-symmetry-protected nodal rings residing on these planes.}
	\label{fig_band}
\end{figure*}

The structural stability of CT-B$_{24}$ is confirmed through multiple criteria. First, the phonon dispersion calculations reveal no imaginary frequencies across the entire Brillouin zone [Fig.~\ref{fig_stability}(a)], confirming dynamical stability. Second, the \textit{ab initio} molecular dynamics simulations performed on a 432-atom supercell at 600~K show stable energy fluctuations around equilibrium over 8~ps, with the structure remaining intact [Fig.~\ref{fig_stability}(b)], demonstrating excellent thermal stability. Third, the mechanical stability is verified via the elastic constants: $C_{11} = 343.42$~GPa, $C_{12} = 108.11$~GPa, $C_{13} = 68.63$~GPa, $C_{33} = 376.12$~GPa, $C_{44} = 181.80$~GPa, and $C_{66} = 117.66$~GPa, which satisfy the Born stability criteria for hexagonal systems~\cite{Born-criteria}: $C_{11} > |C_{12}|$, $2C_{13}^2 < C_{33}(C_{11} + C_{12})$, $C_{44} > 0$, and $C_{66} > 0$. Energetically, the energy per atom is $-6.47$~eV, and a comparison of energy-volume curves with other boron allotropes [Fig.~\ref{fig_stability}(c)] shows that CT-B$_{24}$ is metastable: it lies slightly higher in energy than Pnma-B$_{60}$ (by 0.13~eV/atom) and $\alpha$-B$_{12}$ (by 0.23~eV/atom), but is more favorable than several other predicted phases such as AB-16-Pnnm~\cite{AB-16-Pnnm}, 3D-$\alpha'$ boron~\cite{3DalphaB}, 3D borophene~\cite{3Dborophene}, Ort-B$_{32}$~\cite{Ort-B32}, and H-boron~\cite{H-boron}. The small energy difference relative to $\alpha$-B$_{12}$~\cite{alphaB12}, a near-ground-state stable phase, indicates that CT-B$_{24}$ is synthesizable under nonequilibrium conditions, consistent with numerous metastable functional materials reported previously. Its robust metastability supports its experimental feasibility. A summary of structural and physical properties of CT-B$_{24}$ and other representative boron allotropes is provided in Table~\ref{tab:I} for comparison.

\subsection{Electronic Structure and the Pseudo-Nodal Sphere}

Having established its structural stability, we now examine the electronic properties of CT-B$_{24}$. Since boron is a light element, SOC effects are negligible and were not included in the following analysis. The calculated band structure along high-symmetry paths is shown in Fig.~\ref{fig_band}(a). A striking feature is the crossing between only two bands (the 36th valence band and the 37th conduction band) near the Fermi level ($E_{\text{F}}$), with other bands lying far away in energy. The two bands form Dirac-like linear crossings that are isotropic in all momentum directions. The projected density of states (PDOS) in Fig.~\ref{fig_band}(b) shows that states near $E_{\text{F}}$ are dominated by $p$ orbitals of the six-coordinated boron atoms, consistent with the charge density distribution at the crossing point P in Fig.~\ref{fig_band}(c). A full Brillouin zone scan reveals that these crossings collectively form a quasi-spherical nodal surface centered at $\Gamma$ [Figs.~\ref{fig_band}(j) and ~\ref{fig_band}(k)]. Notably, the density of states near the crossing energy remains nearly constant [Fig.~\ref{fig_band}(d)], matching the theoretical behavior of an ideal NSSM (DOS $\propto \text{const.}$), in contrast to nodal points (DOS $\propto E^2$) or nodal lines (DOS $\propto |E|$).

\begin{figure}[th]
	\centering
	\includegraphics[width=0.48\textwidth]{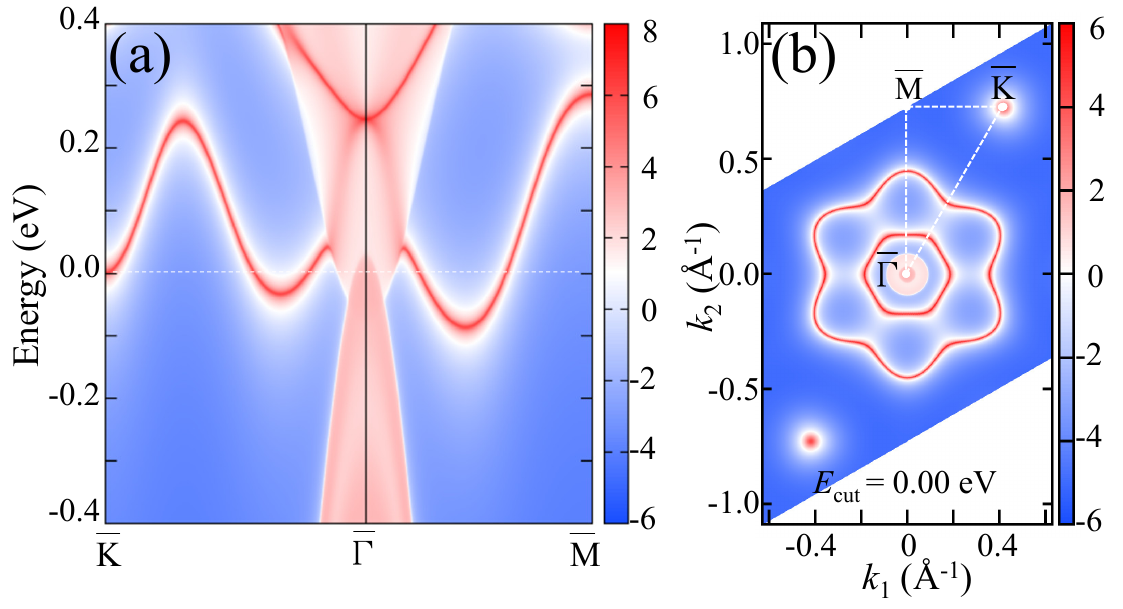}
	\caption{(Color online) Surface states on the (001) surface of CT-B$_{24}$. (a) Surface spectral function along the $\bar{\text{K}}$-$\bar{\Gamma}$-$\bar{\text{M}}$ high-symmetry path. The bright-red intensity denotes topological surface states. (b) Constant-energy contour at the Fermi level ($E_{\text{F}}$), revealing drumhead surface states.}
	\label{fig_surfaces}
\end{figure}

To understand the topological origin of this quasi-nodal sphere, we analyze the symmetry properties of the crossing bands. Along the $\Gamma$--A path, where the bands cross at the D point without gap opening [see Figs.~\ref{fig_band}(e) and~\ref{fig_band}(k)], the little group $C_{6v}$ enforces distinct irreducible representations ($A_1$ and $A_2$) for the two bands, preventing hybridization and ensuring a symmetry-protected crossing. Away from high-symmetry lines, small gaps emerge due to HOTTs, as evident in the band structure [Fig.~\ref{fig_band}(g)]. Remarkably, the maximum gap throughout the BZ is only 0.008~meV, as seen at the $X_1$ point [Fig.~\ref{fig_band}(g)], making CT-B$_{24}$ the closest material realization of an ideal NSSM to date.

\begin{figure*}[!t]
	\centering
	\includegraphics[width=0.76\textwidth]{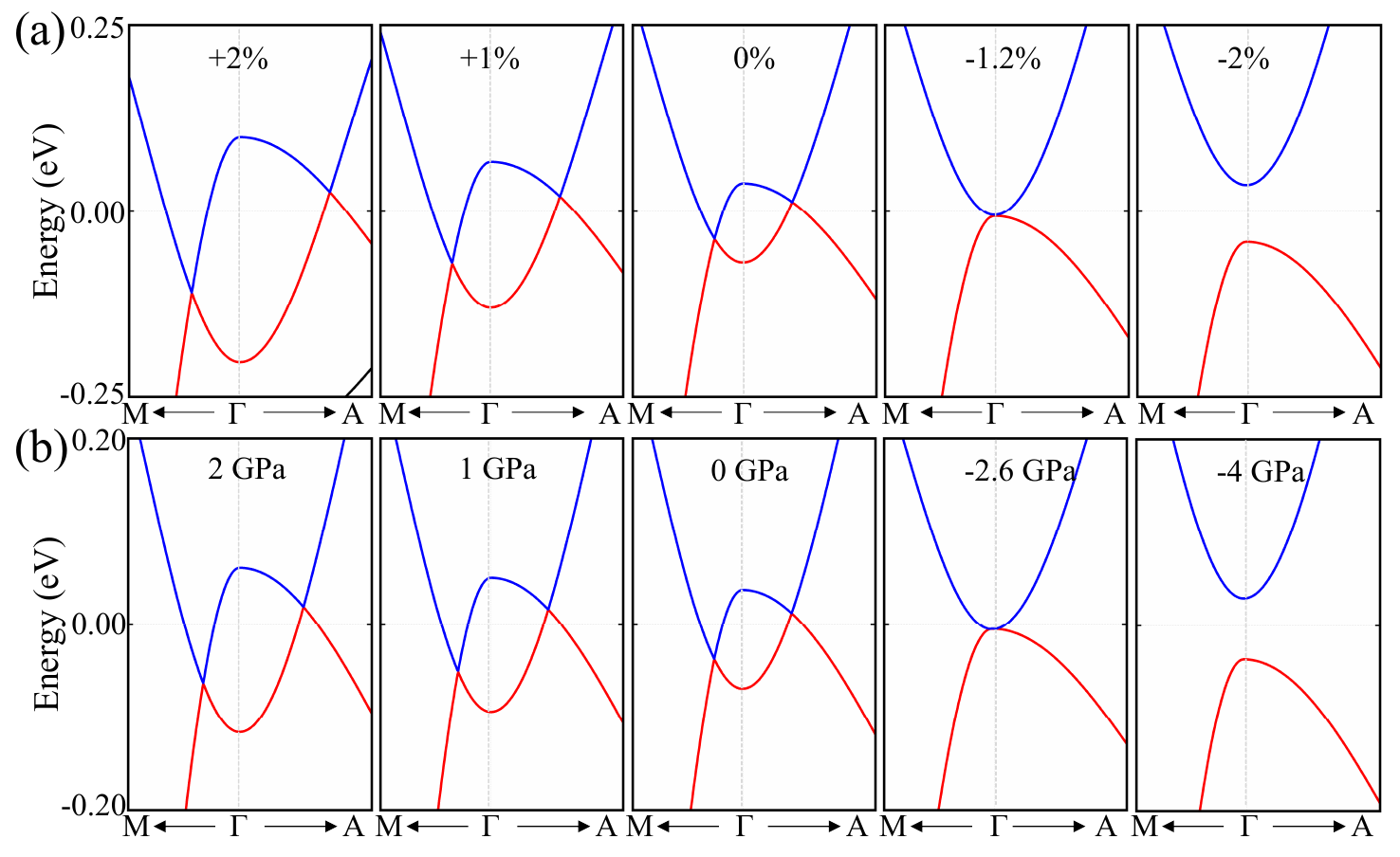}
	\caption{(Color online) Diagram of the band structure evolution of CT-B$_{24}$ under external strain. (a) Uniaxial strain along the $c$-axis. (b) Hydrostatic pressure. The plots trace the topological phase transition from a nodal-sphere semimetal (0\% strain, 0~GPa) to a Dirac semimetal ($-1.2\%$ strain, $-2.6$~GPa) and then to a trivial insulator.}
	\label{fig_strain-bands}
\end{figure*}

The protection by mirror symmetries is confirmed by band eigenvalue analysis [Fig.~\ref{fig_band}(f)], which shows that the crossing bands carry opposite mirror eigenvalues ($\pm 1$), enforcing degeneracy on mirror-invariant planes. Band gap mappings on the $k_x$--$k_z$ and $k_y$--$k_z$ planes [see Figs.~\ref{fig_band}(h) and~\ref{fig_band}(i)] reveal six mirror-protected nodal rings that form a spherical skeleton, as visualized in Figs.~\ref{fig_band}(j) and~\ref{fig_band}(k). Although HOTTs introduce minor gaps at generic $k$-points, the system retains near-degeneracy throughout the nodal sphere, realizing a robust PDNS state. The inset of Fig.~\ref{fig_band}(a) displays a 3D visualization of the band-crossing manifold in the $k_z = 0$ plane, showing a nearly closed ring of crossing points encircling the $\Gamma$ point. Through rotational symmetry, this ring replicates to form a complete spherical manifold in the 3D BZ. Thus, the two bands generate a single PDNS structure rather than isolated nodal rings or nets.

\subsection{Topological Surface States}

To further explore the unique topological characteristics of CT-B$_{24}$, we calculated the surface band structure along the $\bar{\text{K}}$-$\bar{\Gamma}$-$\bar{\text{M}}$ high-symmetry path on the (001) surface projection. On this projected surface, the 3D nodal sphere is mapped onto a 2D nodal ring. As shown in Fig.~\ref{fig_surfaces}(a), we observe prominent drumhead-like surface states that are uniquely nestled outside the projected nodal ring. This characteristic is also evident from the Fermi surface contour of the surface states around the $\bar{\Gamma}$ point in the 2D BZ [Fig.~\ref{fig_surfaces}(b)], which is consistent with previous reports where nodal-sphere surface states are typically confined within or outside the nodal sphere. The presence of these robust surface states provides a crucial experimental signature for the topological nature of CT-B$_{24}$, making it highly amenable to detection via ARPES.

\subsection{Strain-Induced Topological Phase Transition}

The tunability of topological states by external parameters is crucial for both fundamental research and technological applications. We have systematically investigated the effects of $c$-axis uniaxial strain and hydrostatic pressure on the topological state of CT-B$_{24}$. As shown in Fig.~\ref{fig_strain-bands}(a), uniaxial tensile strain along the $c$-axis leads to a continuous increase in the radius of the nodal sphere, indicating a robust response of the topological state to mechanical manipulation. Conversely, compressive strain causes the sphere's radius to shrink. Upon reaching a critical compressive strain of $-1.2\%$, the entire nodal sphere collapses into a single isolated Dirac point at the $\Gamma$ point. Further increasing the compressive strain beyond this critical value opens a global band gap, leading to a topological phase transition from a semimetal to a trivial insulator.

A similar behavior is observed under hydrostatic pressure. Positive pressure increases the radius of the nodal sphere, while negative pressure causes it to shrink, as shown in Fig.~\ref{fig_strain-bands}(b). At a critical negative pressure of $-2.6$~GPa, the nodal sphere undergoes a transition, collapsing to a Dirac point at $\Gamma$ before the system enters a trivial insulating state with a finite gap. The ability to continuously tune the radius of the quasi-spherical degeneracy and reversibly switch the material between a semimetal and a trivial insulator via a simple mechanical action is a powerful finding that could pave the way for the development of novel pressure sensors and strain-engineered quantum electronic devices.

\section{DISCUSSION AND SUMMARY}\label{sec_discussion}

CT-B$_{24}$ represents an exceptionally clean realization of a PDNS semimetal. Its defining feature is the nearly vanishing induced gap ($\approx$0.008~meV). This gap is two orders of magnitude smaller than those in previously proposed PDNS materials. The near-ideal nature of the nodal sphere preserves a constant density of states and isotropic drumhead surface states, leading to pronounced quantum oscillations and distinct plasmonic responses. Symmetry analysis confirms that CT-B$_{24}$ satisfies the criteria for a type-II PDNS semimetal: the two crossing bands possess opposite mirror eigenvalues, and the sixfold rotational symmetry links mirror-protected nodal rings into a continuous spherical degenerate surface, consistent with the 1D irreducible representations of the point group $C_{6v}$.

Structurally, the specific arrangement of B$_3$ triangles and B$_9$ cages in the $P6_3mc$ space-group framework achieves an optimal balance, providing sufficient symmetry to establish the nodal-sphere scaffold while suppressing higher-order coupling terms. The light-element composition of CT-B$_{24}$ ensures negligible SOC, and its drumhead surface states located outside the projected nodal sphere offer clear signatures for ARPES measurements. The robust stability of the structure suggests that it can be synthesized via high-pressure or chemical vapor deposition methods. Furthermore, the nodal sphere exhibits considerable tunability under external strain and pressure; under compressive strains of $-1.2\%$ or hydrostatic pressure of $-2.6$~GPa, it transitions sequentially into a Dirac point and then a trivial insulator. This reversible topological transition suggests potential applications in strain-tunable quantum switches and sensors.

CT-B$_{24}$ is the first boron allotrope reported to host a nodal-sphere state, and it surpasses most topological nodal-line semimetals in boron in terms of band structure cleanliness and energetic metastability. Its spherical Fermi surface and constant density of states open avenues for exploring correlated topological phases, such as unconventional superconductivity and non-Fermi liquid behavior, establishing CT-B$_{24}$ as an ideal material platform for studying 2D band degeneracies and their associated quantum phenomena.

In summary, based on first-principles calculations and symmetry analysis, we predict a new 3D boron allotrope, CT-B$_{24}$, that realizes a nearly ideal pseudo-nodal sphere semimetal state. The stability of CT-B$_{24}$ is thoroughly confirmed by phonon spectra, \textit{ab initio} molecular dynamics simulations, and elastic constant calculations. Our electronic structure analysis reveals that CT-B$_{24}$ exhibits a quasi-spherical band degeneracy near the Fermi level with a record-low maximum gap of only 0.008~meV, making it the closest material realization of an ideal nodal-sphere semimetal reported to date. We show that this quasi-nodal sphere consists of a spherical skeleton of six mirror-symmetry-protected nodal rings spaced $30^\circ$ apart. Projection of the nodal sphere onto the (001) surface reveals pronounced drumhead-like surface states, offering distinct experimental signatures. Furthermore, the quasi-nodal sphere in CT-B$_{24}$ undergoes topological phase transitions under hydrostatic pressure or uniaxial strain along the $c$-axis, evolving from a nodal-sphere state to a Dirac semimetal and finally to a trivial insulator. Our work not only provides an excellent material platform for exploring the physics of near-ideal nodal spheres but also opens avenues for designing strain-tunable topological quantum devices.

\begin{acknowledgments}

This work was supported by the National Natural Science Foundation of China (Grants No.~12304202), Hebei Natural Science Foundation (Grant No.~A2023203007), Science Research Project of Hebei Education Department (Grant No.~BJK2024085), and Cultivation Project for Basic Research and Innovation of Yanshan University (No.~2022LGZD001).

\end{acknowledgments}

\end{document}